\documentclass{article}[12pt]
\usepackage{amssymb}
\begin{document}
\large

\begin{center}
{\Large \bf  Noncommutative tori, Yang-Mills and string theory
}

\vskip .5in

{\large Anatoly Konechny\\
Rutgers, the State University of New Jersey\\
Piscataway, New Jersey, 08854-8019, USA} \\

anatolyk@physics.rutgers.edu

\vskip .5 in

\begin{abstract}

Noncommutative tori are among  historically the oldest
and by now the most developed examples of noncommutative spaces.
Noncommutative Yang-Mills theory  can be obtained from
string theory. This connection led to a cross-fertilization of research
in physics and mathematics on Yang-Mills theory on noncommutative tori.
One important result stemming from that work is the link between T-duality
 in string theory and Morita equivalence of associative algebras.
In this article we give an overview of the basic results in  differential geometry
of noncommutative tori. Yang-Mills theory on noncommutative tori, the duality
induced by Morita equivalence and its link with the T-duality are discussed.
Noncommutative Nahm transform for instantons is introduced.
\end{abstract}
\end{center}

\section{Noncommutative tori}
\subsection{The algebra of functions}
The basic notions of noncommutative differential geometry were introduced
and illustrated on the example of a two-dimensional noncommutative torus
by A.~Connes in \cite{Connes1}.
To define an algebra of functions on a $d$-dimensional noncommutative torus  consider
 a set of linear generators  $U_{\bf n}$ labelled by
${\bf n} \in {\mathbb Z}^{d}$ - a $d$-dimensional
vector with integral entries.
The multiplication  is defined by the formula
\begin{equation} \label{un}
U_{\bf n}U_{\bf m} = e^{\pi i n_{j}\theta^{jk}m_{k}} U_{\bf n + m}
\end{equation}
where $\theta^{jk}$ is an antisymmetric $d\times d$
matrix, and summation over repeated indices is assumed.
We further extend the multiplication from finite linear combinations to
formal infinite series $\sum _{\bf n} C({\bf n}) U_{\bf n}$ where
the coefficients $C({\bf n})$ tend
to zero faster than any power of $\| {\bf n} \|$.
The resulting algebra constitutes
an algebra of smooth functions on a noncommutative torus and will be denoted
as $T_{\theta}^{d}$. Sometimes for brevity we will omit the dimension label $d$ in
the notation of the algebra.
We introduce an involution $*$ in $T_{\theta}^{d}$ by the rule: $U_{\bf n}^{*} = U_{-\bf n}$.
The elements $U_{\bf n}$ are assumed to be unitary with respect to this involution, i.e.
$U^{*}_{\bf n}U_{\bf n} = U_{-\bf n}U_{\bf n} = 1\equiv U_{\bf 0}$.
One can further introduce a norm
and take an appropriate completion of the involutive algebra $T_{\theta}^{d}$ to obtain
a $C^{*}$-algebra of functions on a noncommutative torus. For our purposes
the norm structure will not be important. A canonically normalized trace on
$T_{\theta}^{d}$ is introduced by specifying
\begin{equation}\label{trace}
{\rm Tr} \, U_{\bf n} = \delta_{{\bf n}, {\bf 0}} \, .
\end{equation}

\subsection{Projective modules}
According to  general approach to noncommutative geometry
finitely generated projective modules over the algebra
of functions are natural analogs of vector bundles. Throughout this article when speaking of a
projective module we will assume a  finitely generated left projective module.

A free module $(T_{\theta}^{d})^{N}$ is equipped with a $T_{\theta}^{d}$-valued Hermitian
inner product $\langle . , .\rangle_{T_{\theta}}$ defined by the formula
\begin{equation}
\langle (a_{1}, \dots , a_{N}), (b_{1},\dots , b_{N})\rangle_{T_{\theta}} =
\sum\limits_{i=1}^{N}a_{i}^{*}b_{i} \, .
\end{equation}
A projective module $E$ is by definition a direct summand in a free module. Thus it inherits
the inner product $\langle . , .\rangle_{T_{\theta}}$.
Consider the endomorphisms of  module $E$, i.e. linear mappings $E\to E$ commuting
with the action of $T_{\theta}^{d}$. These endomorphisms form an associative  unital algebra
denoted ${\rm End}_{T_{\theta}}E$. A decomposition $(T_{\theta}^{d})^{N}=E\oplus E'$
determines an endomorphism $P:(T_{\theta}^{d})^{N}\to (T_{\theta}^{d})^{N}$ that projects
$(T_{\theta}^{d})^{N}$ onto $E$. The algebra ${\rm End}_{T_{\theta}}E$ can then be identified
with a subalgebra in ${\rm Mat}_{N}(T_{-\theta}^{d})$ - the endomorphisms of free module
$(T_{\theta}^{d})^{N}$. The latter one has a canonical trace that is a composition of the matrix
trace with the trace specified in (\ref{trace}). By restriction it gives rise to a canonical
trace ${\rm Tr}$ on ${\rm End}_{T_{\theta}}E$. The same embedding also provides
a canonical involution on $E$ by a composition of the matrix transposition and the involution
$\ast$ on $T_{\theta}^{d}$.

A large class of examples of projective modules over noncommutative tori
constitute the so called
Heisenberg modules. They are constructed as follows.
Let $G$ be a direct sum of ${\mathbb R}^{p}$ and an abelian finitely generated group, and let
$G^{*}$ be  its dual group.
  In the most general situation
$G={\mathbb R}^{p}\times {\mathbb Z}^{q}  \times F$ where $F$ is a finite group. Then
$G^{*}\cong {\mathbb R}^{p}\times T^{q} \times F^{*}$.

Consider a linear space ${\cal S}(G)$ of functions on
$G$ decreasing at infinity faster than any power.
We define operators $U_{(\gamma, \tilde \gamma)}: {\cal S}(G)\to {\cal S}(G)$ labelled by a pair
$(\gamma, \tilde \gamma)\in G\times G^{*}$ acting as follows
\begin{equation}\label{U}
(U_{(\gamma, \tilde \gamma)}f)(x)=\tilde \gamma (x) f(x+ \gamma ) \, .
\end{equation}
One can check that the operators $U_{(\gamma, \tilde \gamma)}$ satisfy the commutation relations
\begin{equation} \label{nt}
U_{(\gamma, \tilde \gamma)}U_{(\mu, \tilde \mu)}=   \tilde \mu (\gamma )\tilde \gamma^{-1} (\mu )
U_{(\mu, \tilde \mu)}U_{(\gamma, \tilde \gamma)}   \, .
\end{equation}
If  $(\gamma, \tilde \gamma)$ run over a $d$-dimensional discrete subgroup  $\Gamma \subset G\times G^{*}$,
$\Gamma \cong  {\mathbb Z}^{d}$
then  formula (\ref{U}) defines a module over a $d$-dimensional noncommutative torus $T_{\theta}^{d}$
with
\begin{equation}\label{cocycle}
exp(2\pi i \theta_{ij}) = \tilde \gamma_{i} (\gamma_{j} )\tilde \gamma_{j}^{-1} (\gamma_{i} )
\end{equation}
for a given basis $(\gamma_{i} , \tilde \gamma_{i})$ of the lattice $\Gamma$.
This module is projective if $\Gamma$ is such that $G\times G^{*}/\Gamma$ is compact.
If that is the case then the projective   $T_{\theta}^{d}$-module at hand is called
  a Heisenberg module and denoted by $E_{\Gamma}$.

Heisenberg modules play a special role.
If the matrix $\theta_{ij}$ is irrational in the sense
that at least one of its entries   is irrational then any projective
module over $T_{\theta}^{d}$ can be represented as a direct sum of Heisenberg modules.
In that sense Heisenberg
modules can be used as  building blocks to construct an arbitrary module.

\subsection{Connections}
Next we would like to define connections on a projective module over $T_{\theta}^{d}$.
To this end  let us first
define a  Lie algebra  of shifts $L_{\theta}$  acting  on  $T_{\theta}^{d}$
  by specifying a basis consisting of  derivations
$\delta_{j}:T_{\theta}^{d}\to T_{\theta}^{d}$, $j=1,\dots , d$
satisfying
\begin{equation} \label{deltaj}
\delta_{j} (U_{\bf n}) = 2\pi i n_{j}U_{\bf n} \, .
\end{equation}
These derivations  span a $d$-dimensional abelian Lie algebra that we denote by $L_{\theta}$.

 A connection on a module $E$ over $T_{\theta}^{d}$ is a set of operators
 $\nabla_{X}:E\to E$, $X\in L_{\theta}$
depending linearly on $X$ and satisfying
\begin{equation}
[\nabla_{X}, U_{\bf n} ] = \delta_{X}(U_{\bf n})
\end{equation}
where $U_{\bf n}$ are operators $E\to E$ representing the corresponding generators of $T_{\theta}^{d}$.
In the standard basis (\ref{deltaj}) this relation reads as
\begin{equation}
[\nabla_{j}, U_{\bf n} ] = 2\pi i n_{j}U_{\bf n} \, .
\end{equation}
The curvature of connection $\nabla_{X}$ defined as a commutator
$F_{XY}=[\nabla_{X},\nabla_{Y}]$ is
an exterior two-form on the adjoint vector space $L_{\theta}^{*}$ with values in
$End_{T_{\theta}^{d}}E$.

\subsection{K-theory. Chern character}

The $K$-groups of a noncommutative torus  coincide  with those for
commutative tori:
$$
K_{0}(T_{\theta}^{d}) \cong {\mathbb Z}^{2^{d-1}}\cong K_{1}(T_{\theta}^{d}) \, .
$$

A Chern character of a projective module $E$ over a
noncommutative torus $T_{\theta}^{d}$ can be defined as
\begin{equation} \label{ncChern}
{\rm ch}(E) = {\rm Tr} \, exp\left(\frac{F}{2\pi i }\right) \in \Lambda^{even}(L^{*}_{\theta})
\end{equation}
where $F$ is the curvature form of a connection on $E$,
$\Lambda^{even}(L^{*}_{\theta})$ is the even part of the exterior algebra of $L_{\theta}^{*}$ and
Tr is the canonical  trace  on  $End_{T_{\theta}^{d}}E$.
This mapping gives rise to  a noncommutative Chern character
\begin{equation} \label{ncch}
{\rm ch} : K_{0}(T_{\theta}^{d}) \to \Lambda^{even}(L^{*}_{\theta}) \, .
\end{equation}
The component ${\rm ch}_{0}(E)={\rm Tr}{\bf 1} \equiv {\rm dim}(E)$ is called the dimension of
module $E$.

A distinctive feature of the noncommutative Chern character  (\ref{ncch})
 is that its image does not
consist of  integral
elements, i.e.  there is no lattice in $L_{\theta}^{*}$ that generates the
image of the Chern character.
However there is a different   integrality statement that replaces the  commutative one.
Consider a basis in $L_{\theta}^{*}$ in which the derivations corresponding to basis
elements satisfy (\ref{deltaj}). Denote the exterior forms corresponding to the basis
elements by $\alpha^{1}, \dots , \alpha^{d}$. Then an arbitrary element of
$\Lambda(L_{\theta}^{*})$ can be represented as a polynomial in anticommuting variables
$\alpha^{i}$.
Next let us consider a subset
$\Lambda^{even} ({\mathbb Z}^{d})\subset \Lambda^{even}(L^{*}_{\theta})$ that consists
of  polynomials  in $\alpha^{j}$ having integer coefficients.
It was proved by Elliott  that  the Chern character is injective and its range
 on $K_{0}(T_{\theta}^{d})$
is given by the image of $\Lambda^{even} ({\mathbb Z}^{d})$  under the action of the  operator
$ exp\left(
  -\frac{1}{2}\frac{\partial}{\partial \alpha^{j}} \theta^{jk} \frac{\partial}{\partial \alpha^{k}} \right) $.
 This fact implies that the K-group $K_{0}(T_{\theta}^{d})$ can be identified with the
 additive group  $\Lambda^{even} ({\mathbb Z}^{d})$.

  A K-theory class  $\mu(E)\in  \Lambda^{even} ({\mathbb Z}^{d})$ of a module $E$ can be computed
from its Chern character  by the formula
\begin{equation} \label{Elliott}
\mu(E) = exp\left(
  \frac{1}{2}\frac{\partial}{\partial \alpha^{j}} \theta^{jk}
  \frac{\partial}{\partial \alpha^{k}} \right)
{\rm ch}(E)  \, .
\end{equation}
Note that the anticommuting variables $\alpha^{i}$ and the derivatives
$\frac{\partial}{\partial \alpha^{j}}$ satisfy the
anticommutation relation $\{ \alpha^{i}, \frac{\partial}{\partial \alpha^{j}} \} =
 \delta^{i}_{j}$.

The coefficients of $\mu(E)$ standing at monomials in $\alpha^{i}$ are
 integers  to which we will refer
 as the topological numbers of module $E$. These numbers also can be interpreted
 as numbers of D-branes of a definite kind
although in  noncommutative geometry it is difficult to talk about branes
as geometrical objects wrapped on torus cycles.

One can show that for noncommutative tori $T_{\theta}^{d}$ with irrational matrix $\theta_{ij}$
the set of elements of $K_{0}(T_{\theta}^{d})$ that represent a projective module (i.e.
the positive cone) consists exactly of the elements with a positive dimension. Moreover
if  $\theta_{ij}$ is irrational any two projective modules which represent the same element of
$K_{0}(T_{\theta}^{d})$ are isomorphic that is the projective modules are essentially
specified in this case by their topological numbers.

Complex defferential geometry of noncommutative tori and its relation with mirror symmetry
is discussed in \cite{PS}.

\section{Yang-Mills theory on noncommutative tori}
Let $E$ be a projective module over $T_{\theta}^{d}$.  We call a Yang-Mills field
on $E$ a  connection $\nabla_{X}$ compatible with the Hermitian structure,
 that is a connection satisfying
\begin{equation}\label{herm}
<\nabla_{X}\xi , \eta>_{T_{\theta}} +  <\xi, \nabla_{X}\eta>_{T_{\theta}} =
\delta_{X}(<\xi,\eta >_{T_{\theta}})
\end{equation}
for any two elements $\xi, \eta \in E$.
Given a positive-definite metric on the Lie algebra $L_{\theta}$ we can define a
Yang-Mills functional
\begin{equation}\label{YM}
S_{YM}(\nabla_{i}) = \frac{V}{4g^{2}_{YM}}
g^{ik}g^{jl}{\rm Tr}(F_{ij}F_{kl})\, .
\end{equation}
Here $g^{ij}$ stands for the metric tensor in the canonical basis (\ref{deltaj}),
$V=\sqrt{|{\rm det}\, g|}$, $g_{YM}$ is the Yang-Mills coupling constant, ${\rm Tr}$ stands
for the canonical trace on ${\rm End}_{T_{\theta}}E$ discussed above and
summation over repeated indices is assumed.
Compatibility with the Hermitian structure (\ref{herm}) can be shown to imply
the positive definiteness of the functional $S_{YM}$. The extrema of this functional
are given by the solutions to the Yang-Mills equations
\begin{equation} \label{YMeq}
g^{ki}[\nabla_{k},F_{ij}] = 0 \, .
\end{equation}

A gauge transformation in the noncommutative Yang-Mills theory is specified by a unitary
endomorphism $Z\in {\rm End}_{T_{\theta}}E$, i.e. an endomorphism satisfying
$ZZ^{\ast} = Z^{\ast}Z= 1$. The corresponding gauge  transformation acts on
a Yang-Mills field as
\begin{equation}
\nabla_{j}\mapsto Z\nabla_{j}Z^{\ast} \, .
\end{equation}
The Yang-Mills functional (\ref{YM}) and the Yang-Mills equations
(\ref{YMeq}) are invariant under these transformations.

It is easy to see that Yang-Mills fields whose curvature is a scalar operator,
i.e. $[\nabla_{i}, \nabla_{j}] = \sigma_{ij}\cdot {\bf 1}$ with $\sigma_{ij}$ a
real number valued tensor, solve the Yang-Mills equations  (\ref{YMeq}).
A  characterization of  modules admitting  a constant curvature connection and a
 description of the moduli spaces of constant curvature
connections (that is the space of such connections modulo gauge transformations) is reviewed
  in \cite{KonSch}. Another interesting class of solutions to the
Yang-Mills equations is  instantons (see below).

As in the ordinary field theory one can construct various extensions of the noncommutative
Yang-Mills theory (\ref{YM})  by adding other fields.
To obtain a supersymmetric extension of (\ref{YM}) one needs to add a number of
endomorphisms $X_{I}\in {\rm End}_{T_{\theta}}E$  that play the role of bosonic scalar
fields in the adjoint representation of the gauge group
and a number of odd Grassmann parity endomorphisms
$\psi^{\alpha}_{i}\in \Pi{\rm End}_{T_{\theta}}E$ endowed with an $SO(d)$-spinor index
$\alpha$. The latter ones are analogs of the usual fermionic fields.

In string theory one considers
a maximally supersymmetric extension of the Yang-Mills theory (\ref{YM}).
In this case the supersymmetric action depends on $10-d$ bosonic scalars $X_{I}$,
$I=d, \dots 9$
and the fermionic fields can be collected into an $SO(9,1)$ Majorana-Weyl spinor multiplet
$\psi^{\alpha}$, $\alpha=1, \dots 16$. The maximally supersymmetric
 Yang-Mills action takes the form
\begin{eqnarray} \label{sYM}
S_{SYM} = &&\frac{V}{4g^{2}}{\rm Tr}\Bigl(
F_{\mu\nu}F^{\mu\nu} + [\nabla_{\mu},X_{I}][\nabla^{\mu}, X^{I}]  +
[X_{I}, X_{J}][X^{I}, X^{J}]  \nonumber \\
&&-2\psi^{\alpha}\sigma^{\mu}_{\alpha\beta}
[\nabla_{\mu}, \psi^{\beta}] - 2\psi^{\alpha}\sigma^{I}_{\alpha\beta}
[X_{I}, \psi^{\beta}]\Bigr)\, .
\end{eqnarray}
Here the curvature indices $F_{\mu\nu}$, $\mu,\nu=0, \dots , d-1$
are assumed to be contracted with a Minkowski
signature metric, $\sigma_{\alpha\beta}^{A}$ are blocks of the ten-dimensional
$32\times 32$ Gamma-matrices
$$
\Gamma_{A} = \left(
\begin{array}{cc}
0 & \sigma^{\alpha \beta}_{A} \\
(\sigma_{A})_{\alpha \beta} & 0
\end{array} \right) \, , \qquad A=0,\dots , 9\, .
$$
This action is invariant under
 two kinds of supersymmetry transformations  denoted  by $\delta_{\epsilon}$,
 $\tilde \delta_{\epsilon}$
and defined as
\begin{eqnarray} \label{ncSUSY}
&& \delta_{\epsilon} \psi = \frac{1}{2}(\sigma^{jk} F_{jk} \epsilon + \sigma^{jI}[\nabla_{j}, X_{I}]\epsilon +
\sigma^{IJ}[X_{I}, X_{J}] \epsilon )  \, , \nonumber \\
&& \delta_{\epsilon}  \nabla_{j} =  \epsilon \sigma_{j} \psi \, , \quad
 \delta_{\epsilon} X_{J} =   \epsilon \sigma_{J} \psi \, , \nonumber \\
&& \tilde \delta_{\epsilon} \psi = \epsilon \, , \quad \tilde \delta_{\epsilon}  \nabla_{j} = 0 \, , \quad
 \tilde \delta_{\epsilon} X_{J} = 0 \, .
\end{eqnarray}
where $\epsilon$  is a  constant $16$-component Majorana-Weyl spinor.
Of particular interest for string theory applications are solutions to the equations of
motion corresponding to (\ref{sYM}) that are invariant under some of the above supersymmetry
transformations. Further discussion can be found in \cite{KonSch}.

\section{Morita equivalence}
The role of Morita equivalence as a duality transformation in noncommutative
Yang-Mills theory was elucidated by A.~Schwarz  in \cite{ASMorita}.
We will adopt a definition of Morita equivalence for noncommutative tori
 which  can be shown to be essentially
 equivalent to the standard definition of strong Morita equivalence.
We will say that two noncommutative tori $T_{ \theta}^{d}$ and  $T_{\hat \theta}^{d}$
are Morita equivalent if there exists an $(T_{ \theta}^{d}, T_{\hat \theta}^{d})$-bimodule
$Q$ and an  $(T_{ \hat \theta}^{d}, T_{ \theta}^{d})$-bimodule
$P$ such that
\begin{equation}
Q\otimes_{ T_{\hat \theta}} P \cong T_{\theta} \, , \quad
P\otimes_{ T_{ \theta}} Q \cong T_{\hat \theta}
\end{equation}
where $T_{\theta}$ on the right hand side is considered as a $(T_{\theta}, T_{\theta})$-bimodule
and analogously for $T_{\hat \theta}$.
(It is assumed  that the isomorphisms are canonical).
Given a $T_{ \theta}$-module $E$ one  obtains a $T_{\hat \theta}$-module $\hat E$ as
\begin{equation} \label{mod_map}
 \hat E=P\otimes_{T_{\theta}}E \, .
\end{equation}
One can show that this mapping is functorial. Moreover the bimodule $Q$ provides us
with an inverse mapping  $Q\otimes_{T_{\hat \theta}}\hat E\cong E$.

We further introduce a notion of gauge Morita equivalence (originally called
``complete Morita equivalence'')
that allows one to transport connections along with the
mapping of modules  (\ref{mod_map}).
Let $L$ be a $d$-dimensional
commutative Lie algebra.
We say that $(T_{\hat \theta}^{d},T_{ \theta}^{d})$ Morita equivalence bimodule $P$
establishes a gauge Morita equivalence if it
is endowed with operators $\nabla^{P}_{X}$, $X\in L$ that determine a constant curvature connection
simultaneously with respect to $T_{\theta}^{d}$ and $T_{\hat \theta}^{d}$, i.e. satisfy
\begin{eqnarray}\label{biconnect}
&&\nabla^{P}_{X}(ea)=(\nabla^{P}_{X}e)a + e(\delta_{X}a) \, , \nonumber \\
&&\nabla^{P}_{X}(\hat ae )=\hat a(\nabla^{P}_{X}e) + (\hat \delta_{X} \hat a)e \, , \nonumber \\
&&[\nabla^{P}_{X},\nabla^{P}_{Y}]=2\pi i \sigma_{XY}\cdot {\bf 1} \, .
\end{eqnarray}
Here  $\delta_{X}$ and $\hat \delta_{X}$ are  standard derivations on
$T_{\theta}$ and $T_{\hat \theta}$ respectively.
In other words we have two Lie algebra homomorphisms
\begin{equation} \label{deltas}
\delta : L \to L_{\theta} \, , \qquad \hat \delta : L \to L_{\hat \theta} \, .
\end{equation}

If a pair $(P, \nabla^{P}_{X})$ specifies a gauge  $(T_{\theta},T_{\hat \theta})$
equivalence  bimodule then there exists a correspondence
between connections on $E$ and connections on $\hat E$.
A connection $\hat \nabla_{X}$ on $\hat E$ corresponding to a given connection $\nabla_{X}$ on $E$  is defined as
\begin{equation}\label{conmap}
\nabla_{X} \mapsto \hat \nabla_{X} = 1\otimes \nabla_{X} +  \nabla_{X}^{P}\otimes 1 \, .
\end{equation}
More precisely, an operator
$1\otimes \nabla_{X} +  \nabla_{X}^{P}\otimes 1$ on $P\otimes_{\mathbb C}E$ descends to a
connection $\hat \nabla_{X}$ on $\hat E = P\otimes_{T_{\theta}}E  $.
It is straightforward  to check that under this mapping gauge equivalent connections
go to gauge equivalent ones
$$
\widehat{Z^{\dagger}\nabla_{X} Z} = \hat Z^{\dagger} \hat \nabla_{X} \hat Z
$$
where $\hat Z = 1\otimes Z$ is the endomorphism of $\hat E= P\otimes_{T_{\theta}}E$ corresponding
to $Z\in End_{T_{\theta}^{d}}E$.

The curvatures of
$\hat \nabla_{X}$ and $\nabla_{X}$ are connected by the formula
\begin{equation} \label{curv_shift}
F_{XY}^{\hat \nabla}=\hat F_{XY}^{\nabla} + {\bf 1}\sigma_{XY}
\end{equation}
that in particular shows that constant curvature connections go to constant curvature ones.

Since noncommutative tori are labelled by an antisymmetric $d\times d$ matrix $\theta$,
gauge Morita equivalence establishes an equivalence relation on the set of such
matrices. To describe this equivalence relation
consider an action  $\theta \mapsto h\theta = \hat \theta$   of $SO(d,d|{\mathbb Z})$ on
the space of  antisymmetric $d\times d$ matrices by the formula
\begin{equation} \label{action}
\hat \theta = (M\theta + N)(R\theta + S)^{-1}
\end{equation}
where $d\times d$ matrices $M$, $N$, $R$, $S$ are such that the matrix
\begin{equation} \label{g1}
h= \left( \begin{array}{cc}
M&N\\
R&S\\
\end{array} \right)
\end{equation}
belongs to the group $SO(d,d|{\mathbb Z})$.
The above action is defined whenever the matrix $A\equiv R\theta + S$ is invertible.
One can prove  that two noncommutative tori
$T^{d}_{\theta}$ and $T_{\hat \theta}$ are gauge Morita equivalent if and only if
the matrices $\theta$ and $\hat \theta$ belong to the same orbit of
$SO(d,d|{\mathbb Z})$-action (\ref{action}).

The duality group $SO(d,d|{\mathbb Z})$ also acts on the topological numbers of moduli
$\mu \in \Lambda^{even}({\mathbb Z^{d}})$. This action can be shown to be given by a spinor
representation constructed as follows. First note that the operators $a^{i}=\alpha^{i}$,
$b_{i}=\partial/\partial\alpha^{i}$ act on $\Lambda({\mathbb R}^{d})$ and give
a representation of the Clifford algebra specified by the metric with signature $(d,d)$.
The group $O(d,d|{\mathbb C})$ thus can be regarded as a group of automorphisms acting
on the Clifford algebra generated by $a^{i}$, $b_{j}$. Denote the latter action by $W_{h}$
for $h\in O(d,d|{\mathbb C})$. One defines a projective action $V_{h}$ of  $O(d,d|{\mathbb C})$
on $\Lambda({\mathbb R}^{d})$ according to
$$
V_{h}a^{i}V_{h}^{-1}=W_{h^{-1}}(a^{i})\, , \qquad
V_{h}b_{j}V_{h}^{-1}=W_{h^{-1}}(b_{j}) \, .
$$
This projective action can be restricted to yield a double-valued spinor representation
of $SO(d,d|{\mathbb C})$ on $\Lambda({\mathbb R}^{d})$ by choosing a suitable bilinear form
on $\Lambda({\mathbb R}^{d})$. The restriction of this representation to
the subgroup $SO(d,d|{\mathbb Z})$ acting on $\Lambda^{even}({\mathbb Z}^{d})$ gives
the action of Morita equivalence on the topological numbers of moduli.

The mapping (\ref{conmap})  preserves  the Yang-Mills equations of motion (\ref{YMeq}).
 Moreover, one can define a modification of the Yang-Mills action functional (\ref{YM})
 in such a way
that the values of functionals on $\nabla_{X}$ and $\hat \nabla_{X}$ coincide up to an
appropriate rescaling of coupling
constants. The modified action functional has the form
\begin{equation} \label{modifYM}
S_{YM} = \frac{V}{4g^{2}}{\rm Tr}
(F_{jk} + \Phi_{jk}\cdot {\bf 1}) (F^{jk} + \Phi^{jk}\cdot {\bf 1})
\end{equation}
where $\Phi^{jk}$ is a number valued tensor that can be thought of as some background field.
Adding this term will allow us to compensate for the curvature shift
 by adopting the transformation rule
$$
\Phi_{XY} \mapsto \Phi_{XY} - \sigma_{XY} \, .
$$
Note that the new action functional (\ref{modifYM}) has the same equations of motion
 (\ref{YMeq}) as the original one.

To show that the  functional (\ref{modifYM})
 is invariant under gauge Morita equivalence
one has to take into account two more effects.
Firstly, the values of  trace  change by a factor $c = {\rm dim}(\hat E) ({\rm dim}(E ))^{-1}$
as  $ \hat {\rm Tr} \hat X = c {\rm Tr} X $. Secondly,
the identification of $L_{\theta}$ and $L_{\hat \theta}$
is established by means of some linear transformation $A_{j}^{k}$
the  determinant of which will
rescale the volume $V$.
Both effects can be absorbed into an appropriate rescaling of the coupling constant.

One can  show
 that the curvature tensor, the metric tensor, the background field $\Phi_{ij}$
 and the volume element $V$ transform according to
\begin{eqnarray} \label{tr_rules}
&& F_{ij}^{\hat \nabla} = A^{k}_{i}F_{kl}^{\nabla} A^{l}_{j} + \sigma_{ij} \, , \qquad \hat g_{ij} =  A^{k}_{i}g_{kl}A^{l}_{j}  \, , \nonumber \\
&& \hat \Phi_{ij} =  A^{k}_{i}\Phi_{kl}  A^{l}_{j} - \sigma_{ij} \, , \qquad
 \hat V = V|{\rm det}\, A| \,
\end{eqnarray}
where $A=R\theta + S$ and $\sigma=-RA^{t}$.
The action functional (\ref{modifYM}) is invariant under the gauge Morita equivalence
if the coupling constant transforms according to
\begin{equation} \label{ccMorita}
\hat g^{2}_{YM} = g^{2}_{YM}|{\rm det}\, A|^{1/2} \, .
\end{equation}

Supersymmetric extensions of Yang-Mills theory on noncommutative tori were
shown to arise within string theory essentially in two situations.
In the first case one considers compactifications of the (BFSS or IKKT)
Matrix model of M-theory \cite{CDS}. A discussion regarding the connection
between T-duality and Morita equivalence in this case can be found in section 7 of
\cite{SeibWitt}.
Noncommutative gauge theories on tori can be also obtained by taking the so called
Seiberg-Witten zero slope limit in the presence of a Neveu-Schwarz $B$-field background
\cite{SeibWitt}. The emergence of noncommutative geometry in this limit
is  discussed in the article {\it ``Noncommutative geometry from string theory''} in this volume.
Below we give some  details on the relation between T-duality and Morita equivalence
in this approach. Consider a number of Dp-branes wrapped on $T^{p}$ parameterized by coordinates
$x^{i} \sim x^{i} + 2\pi r$
with a closed string metric $G_{ij}$
and a $B$-field $B_{ij}$. The $SO(p,p|{\mathbb Z})$ T-duality group is represented by
the matrices
\begin{equation} \label{g2}
T= \left( \begin{array}{cc}
a&b\\
c&d\\
\end{array} \right)
\end{equation}
that act on the matrix
$$
E=\frac{r^{2}}{\alpha'}(G + 2\pi \alpha'B)
$$
by a fractional transformation
\begin{equation}\label{Td}
T: E \mapsto E'=(aE + b)(cE + d)^{-1} \, .
\end{equation}
The transformed metric and $B$-field are obtained by taking respectively the symmetric and
antisymmetric parts of $E'$.
The string coupling constant is transformed as
\begin{equation}\label{couplc}
T: g_{s}\mapsto g_{s}'=\frac{g_{s}}{({\rm det}(cE + d))^{1/2}}\, .
\end{equation}

The zero slope limit of Seiberg and Witten
is obtained by taking
\begin{equation}\label{SWlimit}
\alpha' \sim \sqrt{\epsilon} \to 0 \, , \qquad
G_{ij} \sim \epsilon \to 0 \, .
\end{equation}
Sending the closed string metric to zero implies that the $B$-field dominates in the open string
boundary conditions.
In the limit (\ref{SWlimit})
the compactification is parameterized in terms of open string
moduli
\begin{equation}\label{openm}
g_{ij} = -(2\pi \alpha')^{2}(BG^{-1}B)_{ij} \, , \quad
\theta^{ij} = \frac{1}{2\pi r^{2}}(B^{-1})^{ij} \, .
\end{equation}
which remain finite. One can demonstrate  that $\theta^{ij}$ is a noncommutativity
parameter  for the torus and the low energy effective
theory living on the $Dp$-brane is a noncomutative maximally supersymmetric  gauge theory  with
a coupling constant
\begin{equation}
G_{s}=g_{s}\left(\frac{{\rm det}\, g}{{\rm det}G} \right)^{1/4} \, .
\end{equation}
From the transformation law (\ref{Td}) it is not hard to derive the transformation
rules for the moduli (\ref{openm}) in the limit (\ref{SWlimit})
\begin{eqnarray}\label{Tdd}
&&T:g\mapsto g'=(a + b\theta)g(a + b\theta)^{t} \, , \nonumber \\
&&T:\theta \mapsto  \theta' = (c + d\theta)(a + b\theta)^{t} \, .
\end{eqnarray}
Furthermore the effective gauge theory becomes a noncommutative Yang-Mills theory (\ref{sYM})
with a coupling constant
$$
(g_{YM})^{-2}=\frac{(\alpha')^{\frac{3-p}{2}}}{(2\pi)^{p-2}G_{s}}
$$
which goes to a finite limit under (\ref{SWlimit}) provided one simultaneously
scales $g_{s}$ with $\epsilon$ as
$$
g_{s}\sim \epsilon^{(3 - p + k)/4}
$$
where $k$ is the rank of $B_{ij}$.
The limiting coupling constant $g_{YM}$ transforms under the T-duality (\ref{Td}),
(\ref{couplc}) as
\begin{equation} \label{couplec2}
T:g_{YM} \mapsto g'_{YM}=g_{YM}({\rm det}(a + b\theta))^{1/4} \, .
\end{equation}
We see that the transformation laws  (\ref{Td}) and (\ref{couplec2}) have the same
form as the corresponding transformations in (\ref{action}), (\ref{tr_rules}), (\ref{ccMorita})
  provided one identifies  matrix  (\ref{g1}) with  matrix (\ref{g2})
conjugated by $T=\left( \begin{array}{cc}
0&1\\
1&0\\
\end{array} \right)$.
The need for conjugation reflects the fact that
in the BFSS M(atrix) model in the framework of which the Morita equivalence was
originally considered,
the natural degrees of freedom are D0 branes versus
Dp branes considered in the above discussion of T-duality.

One can further check that the gauge field transformations following from gauge
Morita equivalence match with those induced by the T-duality.
It is worth stressing that in the absence of a $B$-field background the effective
action based on the gauge field curvature squared is not invariant under T-duality.

\section{Instantons on noncommutative $T^{4}_{\theta}$}
Consider a Yang-Mills field $\nabla_{X}$ on a projective module $E$ over
a noncommutative four-torus $T_{\theta}^{4}$. Assume that the Lie algebra
of shifts $L_{\theta}$ is equipped with the standard Euclidean metric such
that the metric tensor in the basis (\ref{deltaj}) is given by the identity matrix.
 The Yang-Mills field $\nabla_{i}$  is called an instanton
if the self-dual part of the corresponding curvature tensor is proportional
to the identity operator
\begin{equation}
F_{jk}^{+}\equiv \frac{1}{2}(F_{jk} + \frac{1}{2}\epsilon_{jkmn}F^{mn}) = i\omega_{jk}\cdot
{\bf 1}
\end{equation}
where $\omega_{jk}$ is a constant matrix with real entries. An antiinstanton is defined
the same way by replacing the self-dual part with the antiself-dual one.

One can define a noncommutative analog of Nahm transform for instantons \cite{AstNekSchw}
that has properties very similar to those of the ordinary (commutative) one.
To that end consider a triple $({\cal P}, \nabla_{i}, \hat \nabla_{i})$ consisting of
a (finite projective) $(T_{\theta}^{4}, T_{\hat \theta}^{4})$-bimodule ${\cal P}$,
$T_{\theta}^{4}$-connection $\nabla_{i}$ and $T_{\hat \theta}^{4}$-connection
$\hat \nabla_{i}$ that satisfy the following properties. The connection $\nabla_{i}$
commutes with the $T_{\hat \theta}$-action on ${\cal P}$ and the connection $\hat \nabla_{i}$ with
that of $T_{\theta}$. The commutators $[\nabla_{i}, \nabla_{j}]$,
$[\hat \nabla_{i}, \hat \nabla_{j}]$, $[\nabla_{i}, \hat \nabla_{j}]$
 are  proportional to the identity operator
 \begin{equation}
[\nabla_{i}, \nabla_{j}]=\omega_{ij}\cdot {\bf 1}\, , \quad
[\hat \nabla_{i}, \hat \nabla_{j}]=\hat \omega_{ij}\cdot {\bf 1}\, , \quad
[\nabla_{i}, \hat \nabla_{j}]=\sigma_{ij}\cdot {\bf 1}\, .
 \end{equation}
The above conditions mean that ${\cal P}$ is a $T_{\theta\oplus (-\hat \theta)}^{8}$
module and $\nabla_{i}\oplus \hat \nabla_{i}$ is a constant curvature connection on it.
In addition we assume that the tensor $\sigma_{ij}$ is non-degenerate.

For a connection $\nabla^{E}$ on a right $T_{\theta}^{4}$-module $E$ we
define a Dirac operator $D=\Gamma^{i}(\nabla^{E}_{i} + \nabla_{i})$ acting on
the tensor product
$$
(E\otimes_{T_{\theta}}{\cal P})\otimes S
$$
where $S$ is the $SO(4)$ spinor representation space and $\Gamma^{i}$ are four-dimensional
Dirac gamma-matrices. The space $S$ is ${\mathbb Z}_{2}$-graded: $S=S^{+}\oplus S^{-}$
and $D$ is an odd operator so that we can consider
\begin{eqnarray*}
&&D^{+}: (E\otimes_{T_{\theta}}{\cal P})\otimes S^{+} \to
(E\otimes_{T_{\theta}}{\cal P})\otimes S^{-} \, ,  \\
&&
D^{-}: (E\otimes_{T_{\theta}}{\cal P})\otimes S^{-} \to
(E\otimes_{T_{\theta}}{\cal P})\otimes S^{+} \, .
\end{eqnarray*}

A connection $\nabla^{E}_{i}$ on a $T_{\theta}^{4}$-module $E$ is called
${\cal P}$-irreducible if there exists a bounded inverse to the Laplacian
$$
\Delta = \sum\limits_{i}(\nabla^{E}_{i} + \nabla_{i})(\nabla^{E}_{i} + \nabla_{i}) \, .
$$
One can show that if $\nabla^{E}$ is a ${\cal P}$-irreducible instanton then
${\rm ker}D^{+}=0$ and $D^{-}D^{+}=\Delta$.
Denote by $\hat E$ the closure of the kernel of $D^{-}$. Since $D^{-}$ commutes with
the $T_{\hat \theta}^{4}$ action on $(E\otimes_{T_{\theta}}{\cal P})\otimes S^{-}$
the space $\hat E$ is a right $T_{\hat \theta}^{4}$-module. One can prove that this
module is finite projective. Let  $P:(E\otimes_{T_{\theta}}{\cal P})\otimes S^{-} \to \hat E$
be a Hermitian projector. Denote by
$ \nabla^{\hat E}$ the composition $P\circ \hat \nabla$. One can show that
$ \nabla^{\hat E}$ is a Yang-Mills field on $\hat E$.

The noncommutative Nahm transform of a ${\cal P}$-irreducible instanton
connection $\nabla^{E}$ on $E$ is defined to be the pair $(\hat E,\nabla^{\hat E})$.
One can further show that $\nabla^{\hat E}$ is an instanton.

\vskip .5in

\begin{center}
{\bf \large Acknowledgments}
\end{center}
I am grateful to Albert Schwarz for reading and commenting on the
manuscript of this article.

\vskip .5in
\noindent {\bf \Large Keywords} \\

\noindent BFSS Matrix model, M-theory, Noncommutative geometry, Noncommutative tori,
Noncommutative gauge theory, Morita equivalence, T-duality, Noncommutative instantons,
Noncommutative Nahm transform, Seiberg-Witten decoupling limit


\begin{thebibliography}{99}
\bibitem{Connes1}  Connes A (1980) {\it $C^{*}$ alg\`ebres et g\'eom\'etrie differentielle}.
C. R. Acad. Sci. Paris,
Ser.~A-B, 290. English translation in arXiv:hep-th/0101093.
\bibitem{Connesbook} Connes A (1994) {\it Noncommutative geometry}, Academic Press.
\bibitem{Connesrev} Connes A (2000) {\it Noncommutative geometry year 2000}. math.QA/0011193.
\bibitem{KonSch} Konechny A and Schwarz A (2002) {\it Introduction to M(atrix) theory and
noncommutative geometry}. Phys. Rept. 360,  353-465.
\bibitem{DougNek} Douglas MR and Nekrasov N (2001) {\it Noncommutative Field Theory}.
Rev. Mod. Phys. 73, 977-1029.
\bibitem{Szabo} Szabo RJ  (2003) {\it Quantum field theory on noncommutative spaces}.
Phys. Rep. 378,  207.
\bibitem{RieffelRev} Rieffel MA (1990) {\it Noncommutative tori - a case study of non-commutative
differentiable manifolds}.
Contemp. Math. 105, 191-211.
\bibitem{Rieffel} Rieffel MA (1982) {\it Morita equivalence for operator algebras}. In
{\it Operator algebras and applications}, ed. R.~V.~Kadison, Proc. Symp. Pure Math. {\bf 38}.
Amer. Math. Soc., Providence.
\bibitem{RieffelProj} Rieffel MA (1988) {\it Projective modules over higher-dimensional
non-commutative tori}. Can. J. Math., Vol. XL, No. 2, 257-338.
\bibitem{RS} Rieffel MA and Schwarz A (1999) {\it Morita equivalence of multidimensional
noncommutative tori}. Int. J. of Math.  10 (2), 289.
\bibitem{Li} Li H (2004) {\it Strong Morita equivalence of higher-dimensional noncommutative
tori}. J. Reine Angew. Math.  576, 167-180.
\bibitem{ElliottLi} Elliott GA and Li H (2005) {\it Strong Morita equivalence of higher-dimensional
noncommutative tori. II}. math.OA/0501030.
\bibitem{CDS} Connes A, Douglas MR and Schwarz A  (1998) {\it Noncommutative
geometry and Matrix theory: compactification on tori}. JHEP 02, 003.
\bibitem{SeibWitt}  Seiberg N and  Witten E (1999)
{\it String Theory and Noncommutative Geometry}. JHEP  9909, 032.
\bibitem{ASMorita}  Schwarz A (1998) {\it Morita equivalence and duality}.
 Nucl. Phys.  B534, 720-738.
\bibitem{AstNekSchw}  Astashkevich A,  Nekrasov N and  Schwarz A (2000) {\it On noncommutative
Nahm transform}. Commun. Math. Phys. 211, 167-182.
\bibitem{PS}  Polishchuk A  and   Schwarz A (2003) {\it Categories of holomorphic vector bundles
on noncommutative two-tori}. Commun. Math. Phys. 236, 135-159.
\bibitem{TW} Tang X and Weinstein A (2004) {\it Quantization and Morita equivalence for constant
Dirac structures on tori}. Ann. Inst. Fourier (Grenoble) 54  no. 5,
1565-1580.


\end{thebibliography}
\end{document}